# Subtle interactions for distress regulation: efficiency of a haptic wearable according to personality

Adolphe J. Béquet[a]*, Antonio R. Hidalgo-Muñoz[b], Fabien Moreau[a], Joshua Quick[a], Christophe Jallais[a]

a. *Laboratory Ergonomics and Cognitive Sciences applied to Transport, TS2-LESCOT Univ Gustave Eiffel, IFSTTAR, Univ Lyon, F-69675, Lyon, France*
b. *Institute of Neuroscience of Castilla y León (INCYL), University of Salamanca, Spain*

**Author Note**

Emails: adolphe.bequet@univ-eiffel.fr, arhidalgom@usal.es, fabien.moreau@univ-eiffel.fr, joshua.quick10@gmail.com, christophe.jallais@univ-eiffel.fr

**ABSTRACT**

The incorporation of empathic systems in everyday life draws a lot of attention from society. Specifically, the use of wearables to perform stress regulation is a growing field of research. Among techniques explored, the haptic emulation of lowered physiological signals has been suggested to be promising. However, some discrepancies remain in empirical research focusing on such biofeedback (BF) regarding their efficacy, and the mechanisms underlying the effects of these wearables remains unclear. Moreover, the influence of individual traits on the efficiency of BF has been marginally studied, while it has been shown that personality could impact both stress and its regulation. The aim of this study is to investigate the outcome of interactions with these technologies from a psycho-physiological standpoint, but also to explore whether personality may influence its efficiency when other interaction devices are present. Participants had to play a challenging game while a lowered haptic BF of their heart rate was induced on their wrist. Results showed variable efficiency of the wearable among the participants: a subjective relaxation was evident for the participants exhibiting the highest neurotic and extraverted traits score. Our results highlight the plurality of the modes of action of these techniques, depending on the individual and on the level of stress to regulate. This study also suggests that tailoring these regulation methods to individual characteristics, such as personality traits, is important to consider, and proposes perspectives regarding the investigation of stress and regulation systems embedded in wearables.

*Keywords:* Mindless Computing, Stress Regulation, Real-time Biofeedback, Personality Traits

# 1 INTRODUCTION

Acute stress is a common experience in daily life. It can be a source of motivation or self-confidence. When taking this form, it is referred to as "eustress" and has facilitating effects on activities outcomes. However, it has a counterpart, widely known to the public as "distress" (Healey & Picard, 2005), that can hamper the ability to do activities, or can even lead to long-term adverse effects if experienced too frequently (Chu et al., 2019; Liston et al., 2008). Acute distress can induce behavioral and subjective effects that include attentional tunneling, inefficient decision-making, or discomfort. These effects make acute distress a critical factor for the accomplishment of certain tasks, especially those requiring divided attention. Real-time regulation of acute distress is a growing field in human and affective computing science, with a major challenge: the regulation should not further perturbate the accomplishment of a given task, nor increase distress. The use of wearables seems to be an interesting trail to follow that requires more investigation, especially given that individual factors such as personality traits can influence both stress response and emotional regulation. Some definitions regarding the notions of stress, emotional regulation and personality are introduced below to set the frame of the present work.

## 1.1 Explanatory mechanisms linked to the emergence of acute stress

According to appraisal theories (Scherer, 1999), a context can be evaluated on 2 levels. Primary appraisal consists of an evaluation of the environmental demand of the context. Secondary appraisal is linked to the evaluation of available internal resources to respond to the environmental demand (coping ability). Acute stress emerges when the evaluation of a situation results in a perceived unbalance between environmental demand and the available resources to manage it (Lazarus & Folkmann, 1984). This perceived unbalance between a low appraised coping ability relative to a high appraised situational demand leads to brain and body activity modulations to adapt the organism and restore homeostasis (Herman et al., 2012). Stress response refers to this adaptation process. The nature of subsequent stress (eustress or distress) will depend on the efficiency of the adaptation (Matthews, 2016). The outcome of the adaptation is itself constantly appraised via a feedback loop. To be considered as distressful, stimuli linked to a specific situation (i.e. the stressors) should be appraised as being either novel or unpredictable, with low possibility of control and/or representing a physical or psychological threat (Ferreira, 2019).

## 1.2 Stress as a 3-dimensional response of the cognitive system

The stress response can be considered as a 3-dimensional pattern of activations within the cognitive system. The 3 dimensions involved consist of the neuro-physiological component, linked to brain and body modulations; the subjective (experiential) component, linked to the feelings of the person; and the behavioral component, linked to performance outcome. From a dynamic perspective, these components interact with each other. This definition of stress is illustrated in Figure 1.

**Figure 1**

*Model of stress response (adapted from Cox, 1978)*

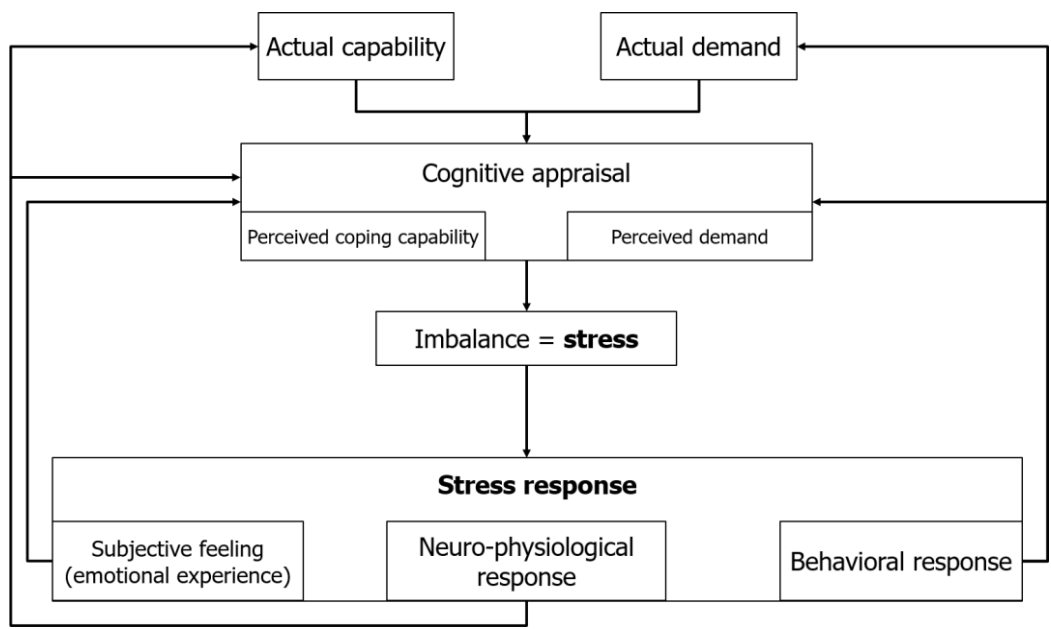

A state of distress is often associated with response patterns involving a high physiological activation (i.e. high arousal), poor performance outcome, and negative feelings such as frustration or impatience (Giannakakis et al., 2019; Helton, 2009).

Specifically, the physiological variations associated with distress have been listed by Giannakakis et al. (2019) in their literature review. The expected variations are: an acceleration of heart rate; a diminution of heart rate variability; an increase in sweat production, reflected by the Electrodermal Activity (EDA) parameters such as the phasic components of the response (number and amplitude of the Skin-Conductance Responses - SCRs) and global level of skin conductance (tonic component); and an increase in respiration rate. According to Giannakakis et al., (2019), and to Benedek & Kaernback (2010), these physiological changes are caused by activation of the sympathetic component of the autonomic nervous system that follows the stress response, driven by brain activations.

### 1.3   Impacts of personality traits on stress and its regulation

Personality traits can influence both stress response and stress regulation efficiency. According to literature, this influence can be exerted through a modulation of attentional mechanisms and coping strategies. There are strong neurological interactions modulating these associations between personality traits, attention, and coping strategies (Van der Linden et al., 2003).

A popular classification of personality traits has been conceived under the term "Big Five" by Goldberg (1981). This classification, based on empirical and lexical observations, describes 5 main personality traits (Plaisant et al., 2010): Openness, referring to a tendency to be curious and insightful; Consciousness, referring to a tendency to be organized and efficient; Extraversion referring to a tendency to be positive and assertive; Agreeability, referring to a tendency to be forgiving and generous; and finally, Neuroticism, referring to a tendency to experience anxiety and tension.

Studies focusing on the link between stress and coping mostly investigated and evidenced a link with neuroticism, extraversion and openness, with mixed results regarding this last trait (Penley & Tomaka, 2002).

Personality traits can be linked to different attentional focuses of individuals regarding environmental stimuli, depending on their valence. Indeed, it has been shown that the neuroticism trait facilitates the treatment of negative stimuli, and makes it harder to disengage attention from them, making neurotic individuals more vulnerable to distress. This could be due to their more important perceived saliency (Bredemeier et al., 2011). Conversely, the extraversion trait is linked to a stronger response towards positive stimuli (Lou et al., 2016). Interestingly, studies report that the subjective impact of stress might also depend on an interaction between the level of stress induction and the personality (Eysenck, 1994): highly neurotic individuals being more exposed in daily stress would have a better tolerance to more intense emotional stress due to a habituation effect (Leblanc, Ducharme & Thompson, 2004).

Beyond attentional bias elicited by personality traits, differences regarding coping strategies have also been found. Neurotic individuals seem more prone to exhibit emotion-focused coping, while extroverts tend to implement problem-focused coping strategies (Penley & Tomaka, 2002). It was also shown that extraverted persons encountering a situation where they have a lack of control might consider this situation as more stressful (LeBlanc, Ducharme & Thompson, 2004).

### 1.4   Toward stress regulation

Focusing on stress regulation, studies tested computer-based biofeedback programs to reduce anxiety (Henriques et al., 2011). However, the regulation strategy tested in this study required a training of 4 weeks to be effective, as well as sustained attention. Moreover, the authors show that these computer-based therapy programs to reduce anxiety require a "series of training sessions and tend to have a relatively high drop-out". Depending on the context in which stress can arise, regulation strategies should interact with users in a subtle manner to ensure that an activity is not disturbed either by stress or by the strategy employed to perform regulation. When technologically implemented, this regulation based on peripheral cues lies in the field of "mindless computing" as defined by Adams et al. (2015). Notably in driving, there has been a rise in studies investigating assistive mindless technologies to regulate emotions (Béquet et al., 2020), but also in office-related environments regarding monitoring of stress induced by computer interactions (Alberti, Aztiria, & Basarab, 2016).

Stress regulation can intervene at different stages of the stress response generation, under various forms. Lazarus & Folkman (1984) identified 2 main coping strategies related to stress regulation: problem-focused coping and emotion-focused coping. The first one consists of acting directly on the source of the stress, while the second one is linked to a modulation of the emotional state itself. These 2 coping strategies echo the model of emotional regulation proposed by Gross (2015). This model identifies 5 stages during which emotional regulation can occur.

The first stage takes place before the emergence of the emotion (here, stress), by means of situation avoidance. Then, during stress generation, situation modification refers to acting directly on the stressor, by suppressing it or by alleviating its stressful component(s). Attentional deployment is a method where the person focuses attention away from the stressors. Cognitive change relates to modulation of the appraisal: it aims at representing the situation differently in order to evaluate it as less stressful. When stress is already strongly developed, the last stage of Gross' model consists of a modulation of the response itself by acting on one of its 3 components (physiological, subjective, behavioral).

The modulation of the physiological stress response can be achieved through an entrainment effect. Entrainment refers to the notion of obtaining a synchronization between an external source and a targeted signal (Clayton, 2012; Phillips-Silver, Aktipis & Bryant, 2010). The purpose of physiological entrainment is to primarily target the physiological component, which could allow a subsequent modulation of the subjective and behavioral components (Trost, Labbé & Grandjean, 2017). Several physiological activities can be targeted to attain this goal. Respiratory activity is an interesting choice, as humans can have voluntary control over it. Thus, this signal can easily be synchronized with an outside stimulus (eg., Lee, Elhaouij & Picard, 2021). Another good candidate is the cardiac response: its entrainment could rely on unconscious processes, which has the advantages of being subtle and not requiring direct attention to be efficient (Azevedo et al., 2017). Moreover, the cardiac component is strongly associated with the feeling of distress and has been exploited for several years to induce discomfort or anxiety, notably through music in the horror film industry (Winters, 2008).

Studies investigating the impact of presenting to participants heartbeat-like stimulations at a lowered rate that their actual heart rate (HR) showed contradictory results. In case of subjective or behavioral modulation, it remains unclear whether these modulations were caused by entrainment or by another process such as attentional deployment. For instance, in his pioneering study, Valins (1966) tricked participants with false heartbeat-like sounds, presenting them as their own. He showed that changes in the subjective evaluation of a visual emotional stimuli could occur, but failed to evidence changes in the actual physiological state. In a replication study, Stern (1972) showed that the subjective effect found by Valins (1996) was rather attributable to various attentional levels toward the auditory stimuli, impacting the attention given to the visual stimuli. Recent studies have shown that interoceptive illusions can be induced by means of false acoustic HR, but without modulating the physiology (Iodice et al., 2019). Taken together, these studies seem to points out an attentional orientation towards the heartbeat-like stimulations, rather than an entrainment effect. However, this conclusion may be tempered, as illustrated by Trost, Labbé & Grandjean (2017) in their review on musically induced entrainment. The authors reported studies where physiological adaptation of the HR to various tempos occurred. The authors attributed this effect to the influence of brain networks linked to the regulation of physiological and emotional processes. Nonetheless, according to the authors, such adaptation may present small effects sizes due to the natural constraints of the cardiovascular system.

Originally, as illustrated above, sounds of heartbeat were mostly used in studies on false HR biofeedback. However, in ecological contexts of stress regulation, the auditory modality may not be the most pertinent one, especially if already solicited by an ongoing task. Indeed, dual-task interferences might arise, especially between 2 task sharing the same (auditory) modality (Wiese & Lee, 2004). Otherwise, tactile stimulation may present several advantages, including avoidance of overload effects that could arise with audio biofeedback. Moreover, haptics can be implemented in various ways and locations over the body (MacLean, 2009), allowing great personalization (Miri et al., 2020), and has been shown to have positive effects on well-being (MacDaniel & Panchanathan, 2020). This modality, unlike sonification (or, for instance, visually-mediated regulation), also presents the advantage of being private (Miri et al., 2020). Finally, tactile stimulation seems to be an efficient way to elicit a representation of physiological responses: studies focusing on cardiac signal have demonstrated that the perception of cardiac activity is linked with mechanoreceptors of the somatosensory system (Couto et al., 2014; Knapp-Kline et al., 2021).

The impacts of wrist-worn haptic biofeedback of the participant's HR on the level of stress have already been tested (Azevedo et al., 2017; Choi & Ishii, 2020; Costa et al., 2017, 2019; Xu et al., 2021) with mixed results. Physiological entrainment of cardiac activity was not systematically found and was associated with small effect size when evidenced. In these studies, vibrations were delivered on a bracelet and at a lowered rate than the actual HR, measured during a resting baseline. Results obtained by these studies are summarized in Table 1. However, in any of the mentioned studies, the lowered HR intended to elicit an entrainment effect was not modulated in real time according to the participants' HR, but systematically used a rate measured during a baseline or fixed at a certain rate. We postulate that a real-time modulation of the lowered rhythm according the participants' real HR may elicit a greater entrainment effect. Moreover, while in their study of haptic false HR biofeedback, Xu et al. (2021) explored the influence of interoceptive accuracy on the efficiency of the regulation, individual characteristics, such as personality traits linked to emotional regulation, have not been investigated.

**Table 1**

*Summary of the studies focusing on a wrist-worn haptic biofeedback of heart rate to reduce stress/anxiety. BF= Biofeedback; BFNE= Brief Fear of Negative Evaluation Scale; EDA= Electrodermal Activity; HBC= Heartbeat Counting; HR= Heart Rate; HRV= Heart Rate Variability; IAC= Interoceptive accuracy*

| Study | Stress induction | Individual traits evaluation? | Real time cardiac stimulation? | Physiological modulation? | Subjective modulation? | Behavioral modulation? | Individual traits impact |
|---|---|---|---|---|---|---|---|
| Azevedo et al., 2017 | Public speech task | Social anxiety (BFNE) | No (baseline 5 mins), frequency 20% lower than HR during baseline | Under stress: smaller increase in physiological arousal (measured using EDA) in the BF group. | Under stress: smaller Increase in anxiety in BF group compared to control group | Not tested | No |
| Costa et al., 2017 | Public speech task | None | No Frequency fixed at 60 bpm for the slow group; Real time for the real HR group | Not tested | Lower anxiety score for the BF group | Not tested | Not tested |
| Costa et al., 2019 | Modular arithmetic problems task | None | No Frequency 30% lower than participant's baseline HR | Higher HRV for slow BF | Lower anxiety for slow BF group | Participants in slow BF group answered more questions correctly, missed less questions, but took more time to respond | Not tested |
| Choi & Ishii, 2020 | Physical stress task (jumping jacks) | None | No Frequency fixed at 60bpm | Bigger decreasing HR and bigger increasing HRV after an exercise for slow BF group | Not tested | Not tested | Not tested |
| Xu et al., 2021 | Trier social stress test | IAC (using HBC) task); Liebowitz social anxiety scale | No (baseline 3mins), frequency at 80% of participant's HR during baseline | Under social stress, with BF, participants with higher IAC showed less increase in HR than participants with lower IAC | No | No | Impact of IAC |

### 1.5 Objectives of the current study

This study examined the impact of a subtle regulation interface, taking the form of a lowered tactile biofeedback of the participant's heart rate. A major novelty regarding previous studies was the adaptation of the frequency of vibrations in real-time, and the context of use of this interface. This regulation took place during an interaction with a touchpad memory game designed to elicit stress, together with a task consisting of the monitoring of auditory alarms. Specifically, we are investigating 1/ whether a reduced real-time biofeedback (BF) of heart rate could elicit a modulation of the physiological component of the stress response. 2/ the influence of BF on the behavioral and subjective components of the stress response. 3/ the influence of personality traits on BF efficiency, that is, on stress regulation.

It is hypothesized that:

- A haptic wearable providing BF will elicit a modulation of the physiology, resulting in a lowered arousal when the BF is present in a stressful context. Specifically, this effect should be reflected by an entrainment of the heart rate, with a drop during exposure to regulation.

-This entrainment effect will lead to subjective and behavioral impacts, resulting in a reduced feeling of distress, and performance improvements, regarding the interaction with a stressful game and alarm detection.

-Personality factors will impact stress and the efficacy of the wearable. Specifically, it is expected that neuroticism will be associated with a stronger distress response, while extraversion should be linked with an increased efficiency of the regulation.

## 2 METHODS

### 2.1 Participants

The participants were 29 healthy French right-handed adults (17 males, 12 females) aged 19-60 years ($M = 34.1$, $SD = 10.9$) recruited through social networks such as Facebook or LinkedIn. All participants had normal auditory acuity and normal, or corrected-to-normal, vision. None of them declared cardiorespiratory or neurological diseases, and were not undergoing any medical treatment that could have impacted their vigilance. All participants were asked to sign a written consent. The study complied with the Declaration of Helsinki for human experimentation and the experimental procedure was approved by the local ethics committee of University Gustave Eiffel. Participants received a financial compensation of €50.

## 2.2 Procedure and experimental design

In order to induce stress, we designed a task consisting of a visual memorization game presenting various levels of difficulty and stress. This task was always concurrent with an auditory detection task, implemented to further explore behavioral impacts of stress and its regulation in a peripheral monitoring context. The biofeedback regulation intervened during the execution of the stressful game.

Participants went through 5 experimental blocks with intermediate surveys between each block, measuring their level of mental workload and their feelings during the previous block. The first block was training during which the dual-task was ran for 3 minutes. The game was set in easy mode for this block. The training surveys were given to the participant, in order to give him the possibility to ask any relevant question before the completion of the actual task survey. The design is summarized in Figure 2.

Each block ran the dual task for 8 minutes, and the questionnaires took less than 5 minutes to complete. The presentation order of the blocks was counter-balanced except for the Easy Game, which was always the first block to be completed. This was done to exploit it as a baseline. The passation order of DG, DSG, DSGR conditions was randomized to avoid fatigue or habituation bias on the data. Participants were wearing the biofeedback device from the beginning of the study, but it was activated only during the corresponding block.

**Figure 2**

*Overview of experimental procedure, with one possible order of passation. DG = Difficult Game; DSG = Difficult Stressful Game; DSGR = Difficult Stressful Game with Regulation*

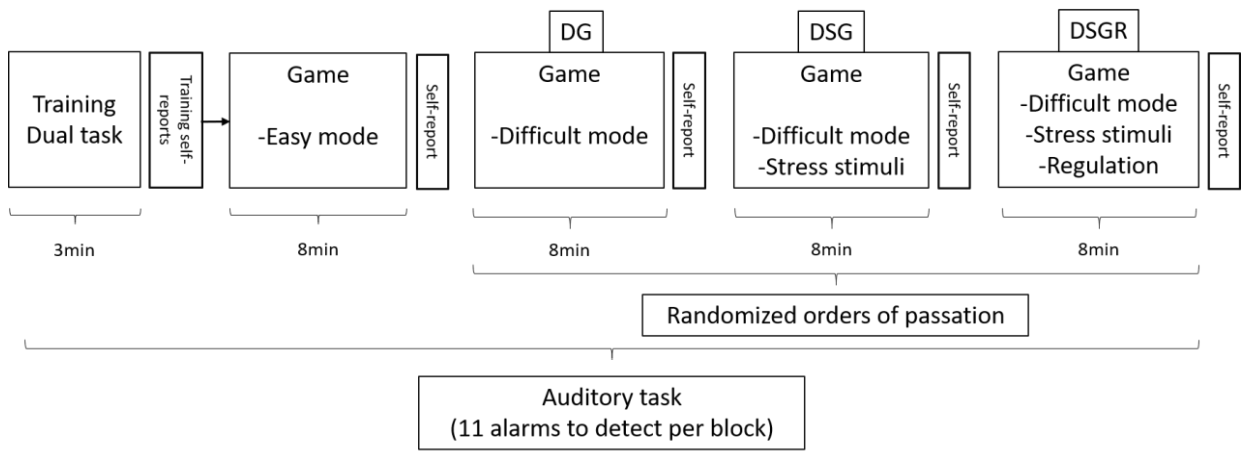

After the completion of the experimentation, a debrief with the participant allowed the collection of additional subjective data: the level of overall distress felt through the experimentation and the associated most distressful elements, the level of distraction induced by the biofeedback, and additional opinions regarding the biofeedback (agreeability, synchronization with the actual heart rate, impact on stress).

### 2.3   Data Collection

Physiological (cardiac, respiratory and electrodermal signals) and behavioral (dual task performance) data was collected and synchronized using the software RTMaps (version 4.6.0).

#### 2.3.1   *Subjective*

Subjective data was collected using a variety of questionnaires. Before the start of the experiment, we measured personality traits using: The French version of the Big-five (Plaisant et al., 2010). Moreover, we measured additional individual factors, such as gender, age, education level, musical and gaming practice.

During the experimentation, self-report questionnaires were delivered at the end of each block: the level of cognitive workload, using the NASA-TLX (Hart & Staveland, 1988), and subjective feelings on positive and negative valences, using the Geneva Emotional Wheel (GEW, Scherer et al., 2005). Feelings displayed on the wheel were inspired from the items of the Short Stress State Questionnaire (Helton et al., 2004) and were the following for GEW positive feelings: *alertness, motivation, self-confidence, serenity, joy*; and for GEW negative feeling: *impatience, frustration, sadness, social worry, dissatisfaction*.

### 2.3.2    *Physiological*

An electrocardiogram (ECG) signal was recorded (sampling rate = 1 kHz, hardware low-pass filtering 35 Hz) throughout each block by placing 3 electrodes on the right clavicle, under the last left rib and on top of the right hip of the participant. The respiration signal (sampling rate = 1 kHz, hardware low-pass filtering 10 Hz) was recorded using a respiratory belt transducer. Both signals were registered via a wireless device (BioNomadix system with MP150 Biopac system). Electrodermal activity (sampling rate = 1 kHz, hardware low-pass filtering 10 Hz) was recorded using 2 electrodes placed on the distal phalange of digits II & III of the non-dominant hand of the participant (Boucsein, 2012, p106). The electrodes used to collect ECG and EDA were disposable BIOPAC pre-gelled electrodes (EL503 for ECG, EL507 for EDA) for which technical details can be found on the constructor's website[1].

### 2.3.3    *Behavioral*

The behavioral data acquisition included reaction times to the auditory detection task (measured using the footswitch), and performance measures of the game described below: length of each pattern composing a trial, time of completion of each trial, and result of the trial. Using these parameters, we created a global indicator (Game performance indicator) for each participant that was normalized across participants.

## 2.4    Dual task implementation

### 2.4.1    *Touchpad "Simon says" game task*

The memorization game was implemented on a MIDI controller, or "touchpad" (Figure 3). The device used was a Novation Launchpad MKII and presented an 8×8 LED button grid with 2 additional rows of LEDs on both sides of the grid. The device offered the possibility to program the game and collect data with great flexibility. Moreover, this format presents the advantage of being ergonomic and very attractive for the participants compared to using a traditional tablet.

The game was programmed using Python and the toolbox "launchpad_py" available on Github[2]. The game requires the participant to memorize and repeat random green light patterns displayed on the grid using the LED buttons. A trial is composed of a red fixation cross at the center of the grid, displayed for 500 milliseconds. Then the pattern was displayed by sequentially lighting up the buttons at a speed of 250 milliseconds per button. The complete pattern then remained displayed for 500 milliseconds before disappearing. The participant then had to reproduce the pattern by sequentially pressing the corresponding buttons to light them up again immediately after their disappearance. In the case of a mistake, the participant could correct it by pressing the button again to turn it off. Finally, the participant had to validate his response by pressing a green button localized on the top left corner of the touchpad. A feedback of the performance was given on the touchpad, taking the shape of a blue circle upon case of success, or a red cross otherwise. Participants were instructed to respond using only their right hand. Figure 3 illustrates the course of a trial.

---

[1] https://www.biopac.com/products/
[2] https://github.com/FMMT666/launchpad.py/tree/master/launchpad_py

There were 2 levels of difficulty, in terms of pattern length. In the easy level, patterns were composed of 1 to 7 buttons to press maximum. In the difficult level, patterns were composed of at least 7 buttons to press, with no limitation of the maximum length. The difficulty adapted itself to the participants within the above-mentioned boundaries: depending on the participant's result from the last 2 trials, the number of buttons to press increased (in the case of 2 consecutive successes) or decreased (in the case of 2 consecutive failures) by 1. This adaptation was made to ensure that the difficulty did not reach a level that would have been too hard for the participant. Furthermore, in order to avoid a disengagement from the task by the participants, the level of the difficulty (in terms of speed of apparition/disappearance) was pre-tested on 5 volunteers.

Stress was manipulated using the additional 8-buttons rows located on both sides of the touchpad as gauges and was inspired by the MIST (Dedovic et al., 2005). 3 manipulations were simultaneously made:

-A temporal pressure, as the higher gauge represented a time constraint. This gauge was fully lit up at the beginning of each trial, and the buttons sequentially turned off during the reproduction of the pattern by the participant. If the participant failed to reproduce the pattern before the complete extinction of the gauge, the trial was considered as failed, and the red cross feedback was displayed. The speed at which the lights went out depended on the participant's performance during the previous trial: in the case of failure, the time constraint was increased by 1000 milliseconds (ms) to keep the difficulty adaptative, in the case of success, the constraint was reduced to the participant's last duration of completion + 100 ms. With this simple implicit rule, the participant was in a self-competitive situation across the trials.

-A performance pressure, with the score of the participant displayed on the lateral gauge, with a label indicating an objective to follow. The participant was instructed that his performance was evaluated, and that he had to reach at least the objective indicated on the gauge.

-A social pressure, induced through an instruction given to the participant. He was told that the score displayed on the gauge was reflecting his score relative to the other participants of the study. Moreover, it was specified that if the participant failed to reach the objective, his data would not be considered in the data analysis due to too much deviation from the other participants.

This performance gauge was also manipulated: in case of trial failure, the score was diminished by 2 points. In case of success, the score was increased by only 1 point.

**Figure 3**

*Presentation of the touchpad game main steps during Difficult Stressful Game condition with a) presentation of a pattern b) lighting up of the temporal gauge, and c) feedback of participant's performance (in case of failure) and update of the performance/objective gauge*

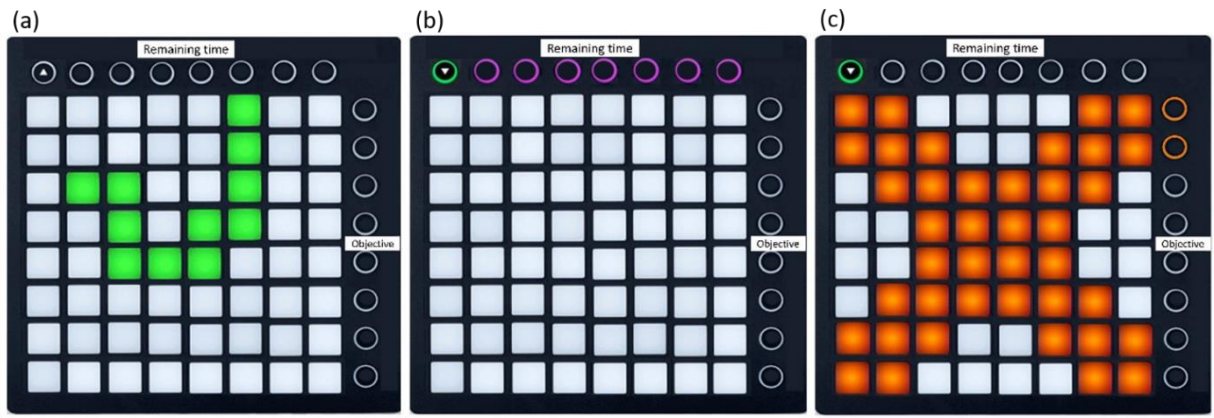

### 2.4.2  Auditory detection task

The auditory task consisted in the detection of shorts bleeps. These sounds, made using the software Audacity (version 3.0), were 100 ms long and had a frequency of 1 KHz. They were presented to the participants at a volume of 75 dB. A background white noise was displayed throughout the task at a volume of 65 dB. Sound intensity was controlled using a sonometer.

Participants were instructed to press a pedal on a footswitch located in front of them "as fast as possible" using their right foot each time they heard the bleeps. This modality of response was chosen because the participant wore the electrodermal electrodes on their left hand, and had to complete the game using their right hand. They also were instructed to keep their foot down near the footswitch between each sound trial, to ensure that all participants were initiating their motor response from the same location. Thus, an auditory omission is defined as the non-response of a participant to an alarm.

## 2.5 Biofeedback implementation

In order to deliver a biofeedback of the heart rate, the collected cardiac signal was processed on the experimental computer in real-time. We detected R-peaks, a component of the cardiac signal that reflects the heartbeat. We developed a MATLAB (version R2020b) module that was implemented in the RTMaps software to perform real-time R-peaks detection, based on their prominence and on inter-peaks distance (Figure 4). The inter-peak distance was then used to determine the real-time heart rate.

**Figure 4**

*Illustration of the real-time detection of cardiac signal's R-peaks, detected during the experimentation. Amplitude is in millivolts (mV).*

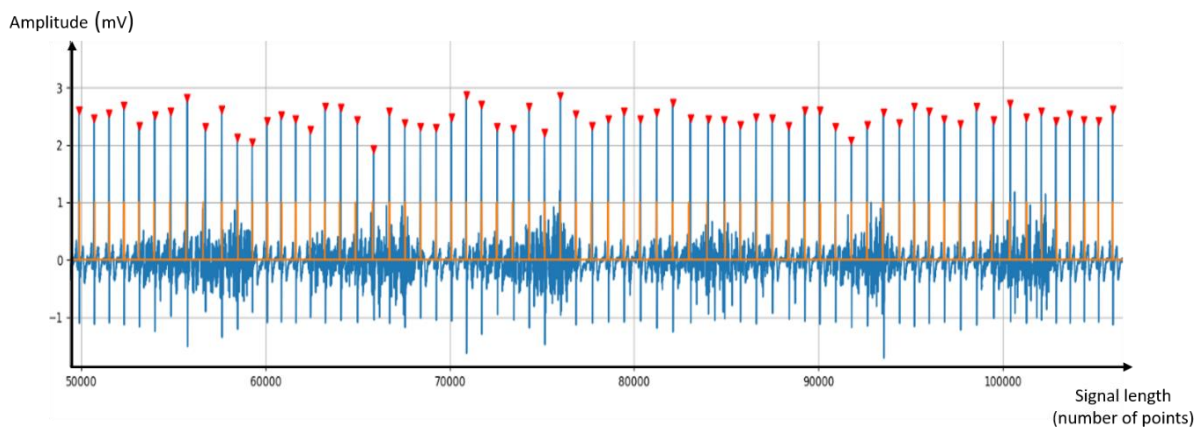

Haptic biofeedback was implemented on a bracelet (Figure 5) inspired by Costa et al. (2016). The bracelet was composed of 3-coin vibrations motors (Precision Microdrives) located on the left inner-wrist of the participant. To communicate with the motors, a micro-controller board (DFRobot Beetle BLE) was included on the top of the bracelet. This board was programmed, using the Arduino software (version 1.8.12), to activate vibration patterns upon the reception of a trigger, emitted from the experimental computer. The vibration patterns were composed of a fast double impulse vibration of the 3 motors, following the recommendations of Miri et al. (2017). According to the authors, such double tapping would elicit a greater association of the delivered vibrations with the heart rate. The intensity of the vibrations was pre-tested on the same 5 volunteers on which the game was pre-tested, to ensure that it was not too distracting nor too imperceptible.

The inter-vibration intervals were set to be larger than the actual real-time measured heartbeat intervals by a factor of 1.5 in order to elicit an entrainment effect on the participant's actual heart rate, following the recommendations of Costa (2019) that the rate delivered should be of about 30% of the actual HR.

**Figure 5**

*Bracelet used to perform biofeedback. The motors are shown on top for visualization purposes. On the actual device, these are located on the inner wrist zone.*

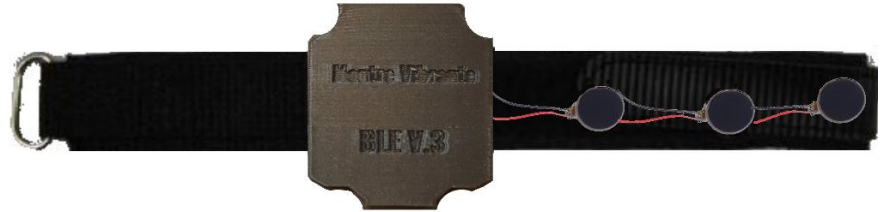

Participants were told that the vibrations reflected their actual heart rate, that there were no specific instruction regarding these vibrations and that they could ignore them.

The device was connected to the computer using a micro USB C-type cable. Triggers were sent using RTMaps in order to start each double impulse vibration. We initially considered using Bluetooth 4.0 to transmit these triggers, but we encountered battery issues, as the device's consumption was too great. The slack of the cable was sufficient to not disturb the participant.

### 2.6 Experimental set-up

The participants were comfortably seated, and were surrounded by 4 loudspeakers. There was a footswitch (Logitech G25) in front of them, at a distance that was decided by the participant to ensure that he had no difficulty reaching the pedal. The game was located in front of them, at a distance of approximately 50 centimeters. The duration of the whole procedure (questionnaires, equipment with data collection, accomplishing the dual task, and debrief) was approximately 2 hours and was composed of a short training and 4 blocks of an 8-minute duration. The dual task itself lasted for about 50 minutes with pauses between blocks to complete mental workload and stress questionnaires. The tasks and data collection were implemented using the experimental software RTMaps. Figure 6 presents the experimental setup.

**Figure 6**

*Technical setup of the experimentation*

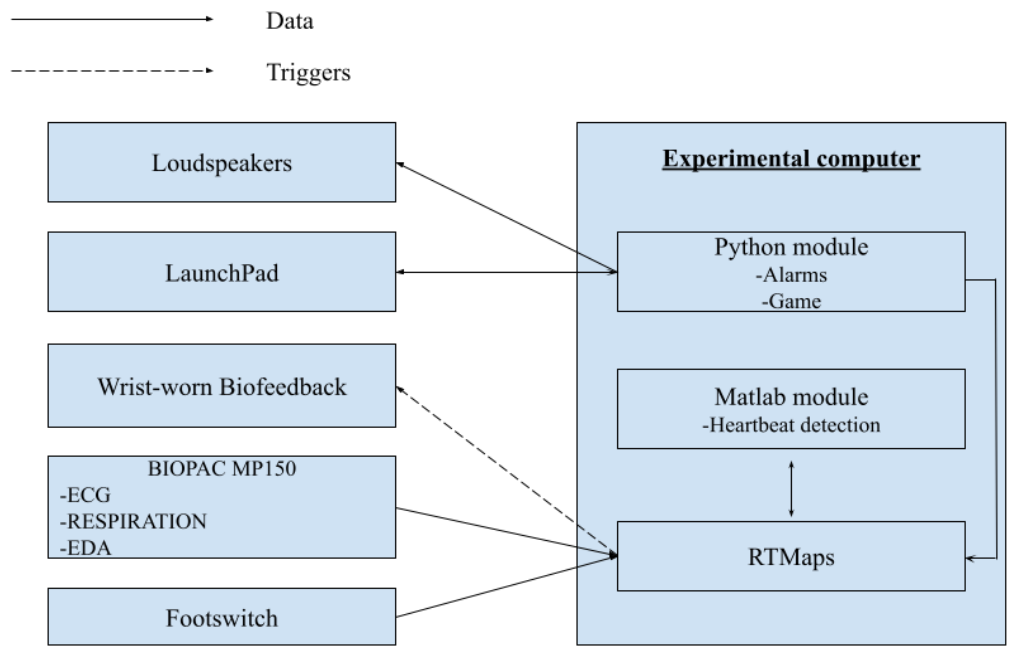

## 2.7 Data processing & cleaning

### 2.7.1 Physiological data

A manual check of signals was conducted by 2 authors to remove signals containing artefacts and noise before features calculation. The manual check performed was specifically looking for important amplitude drifts, noisy signals (containing variations at a frequency superior than the expected frequencies for these data: 0.8-3 Hz for cardiac signal, 0.1-0.8 Hz for respiratory and 0.1-0.3 Hz for electrodermal signals), or signal duration shorter than the 480 seconds of each condition. We then discarded across the conditions 3 participants for cardiac signal, 7 participants for respiration and electrodermal signals.

Physiological features were computed for each condition for the whole duration of the condition (0-480 seconds). Cardiac and respiratory features were computed using dedicated MATLAB scripts. For the cardiac signal, similarly to the real-time detection, we detected R-peaks based on their prominence and inter-peaks distances. Peak detection was visually checked for algorithmical mistakes. We then computed the mean heart rate (HR), and the heart rate variability indicator Root Mean Square of Successive Differences (RMSSD). Respiration Rate (RR) was also computed by detecting inspiration peaks on the filtered respiratory signal. Electrodermal response features were computed using the MATLAB toolbox Ledalab (version 3.4.6, Benedek & Kaernbach, 2010). The EDA signal was decomposed in tonic and phasic components using continuous decomposition analysis. The amplitude threshold of skin conductance phasic responses (SCRs) was set to .015 µS (Wickramasuriya & Faghih, 2019). From these calculations, we obtained the number and amplitudes of SCRs in the EDA signal, as well as the mean amplitude of the tonic component of the EDA signal. We obtained the normalized physiological parameters $P_{norm}$ considering the relative difference according to the baseline (easy game): $P_{norm} = \frac{P_x - P_b}{P_b}$, with $P_x$ being the non-normalized value for parameter $P$ in the condition $x$ (e.g. HR cond. DSG), and $P_b$ the baseline value of this parameter (e.g. HR cond. EG).

### 2.7.2 Behavioral data

Regarding game performance, as there was no possibility of timeout outside DSG and DSGR, in order to compare the performance regarding between the difficult condition and the stressful conditions, we created 2 global indicators linked to correct pattern completion performance and timeout performance.

Regarding alarm response, we collected the amount of footswitch presses that followed an alarm. We eliminated the occurrence of shorts double presses and calculated the reaction time by subtracting the press timestamp from the alarm emission timestamp.

### 2.7.3 Subjective data

The subjective component was evaluated using the global scores on the NASA-TLX scale (reflecting perceived task difficulty out of 120), and the global score of negative feelings (reflecting perceived feeling of distress, on a 5 points scale), measured using the Geneva Emotional Wheel. These questionnaires, as well as the Big-five, were computed using Microsoft Excel.

### 2.7.4 Statistics

Statistical analyses were performed using the open-source software JASP (version 0.16). The significance level was set to $p < .05$. Normality was first checked for the parameter value distributions using the Shapiro-Wilk's test. Cronbach's Alpha was calculated for the GEW negative feelings items, and for the NASA-TLX items, these values were $α > .6$ for each condition, which is acceptable according to Taber (2017). We also calculated the Cronbach's Alpha for the Big five items regarding each personality dimension. The obtained values were $α > .8$, which indicates a high reliability of these questionnaires (Taber, 2017).

We tested that our protocol successfully induced a distress response on the 3 components of the cognitive system. This was done using paired one-tailed *t*-tests (or one-tailed Wilcoxon tests when normality could not be assumed) comparing the features from the Difficult Game (DG) and the Difficult and Stressful Game (DSG) conditions on the 3 dimensions (NASA-TLX; GEW negative feelings score; HR; RMSSD; number and amplitude of SCRs and mean amplitude of the tonic component of the EDA; game performance; number of alarm omissions and detection reaction times). Effect sizes were computed for significant comparisons (Cohen's *d* for *t*-tests, and matched-pairs rank biserial correlation for Wilcoxon tests).

Secondly, we tested whether the real-time biofeedback succeeded in eliciting a physiological entrainment. This was done using paired two-tailed *t*-tests (or Wilcoxon tests) between DSG and Difficult and Stressful Game with Regulation (DSGR) conditions. Using the same comparisons, we also tested whether there was a subjective or behavioral effect of the biofeedback regulation. In order to further explore a potential impact of biofeedback, we constructed 2 groups of participants according to its perceived efficiency. These groups were constituted using the frustration/stress item of the NASA-TLX: we used the median of scores variations, between DSG and DSGR, as a reference to group participants according to their score variation on this item. We then conducted a repeated-measures ANOVA on the GEW negative feelings scores and on physiological parameters, with the biofeedback perceived efficiency groups as between subject factor. We explored the differences between these groups using post-hoc analysis (Bonferroni correction).

Finally, we explored the impact of personality traits on the efficiency of biofeedback using correlation analysis between traits and subjective components (Spearman's correlation). Driven by the analysis' results, groups of participants were formed depending on their extraversion and on their neuroticism scores, on the basis of the median of these scores. The groups were balanced ($N = 15$ participants, $M = 2.17$, $SD = 0.50$, in Low neuroticism group; $N = 14$, $M = 3.72$, $SD = 0.69$, in High neuroticism group; $N = 14$, $M = 2.66$, $SD = 0.30$, in Low extraversion group; $N = 15$, $M = 3.97$, $SD = 0.61$ in High extraversion group). One-way ANOVA with the personality groups between subject factors was then conducted on the features for the 3 stress dimensions.

# 3 RESULTS

This section describes the results obtained from the study. Table 2 reports the means and standard deviations for the subjective, physiological, and behavioral features investigated in the following sections. We report non-normalized values for the physiological component. Regarding the behavioral component, the total number of alarm omission for each condition was 2 for DG; 8 for DSG and DSGR.

**Table 2**

*Mean (standard deviations) of investigated features for each condition. DG = Difficult Game; DSG = Difficult Stressful Game; DSGR = Difficult Stressful Game with Regulation*

| Component | Variable | DG | DSG | DSGR |
| --- | --- | --- | --- | --- |
| Subjective | Global NASA-TLX | 73.2 (16.6) | 87.8 (13.6) | 85.0 (15.5) |
| | Stress subscale of NASA-TLX | 12.3 (4.9) | 14.8 (3.8) | 13.8 (4.3) |
| | Negative feelings on GEW | 2.1 (0.9) | 2.5 (0.8) | 2.3 (0.9) |
| Physiological | HR (beat per min) | 77.56 (9.13) | 80.62 (11.52) | 81.12 (11.19) |
| | RMSSD (ms) | 75.96 (39.92) | 76.84 (49.38) | 80.63 (50.93) |
| | RR (respiration cycles per min) | 18.96 (2.73) | 19.76 (2.52) | 19.86 (2.65) |
| | Number of SCRs | 26.92 (15.96) | 33.77 (16.65) | 37.08 (19.43) |
| | SCRs Amplitude (µSiemens) | 0.04 (0.02) | 0.04 (0.02) | 0.04 (0.01) |
| | Tonic EDA (µSiemens) | 0.64 (0.28) | 0.63 (0.30) | 0.64 (0.29) |
| Behavioral | Game performance indicator | 1.03 (0.21) | 0.99 (0.15) | 0.96 (0.17) |
| | Alarm detection reaction time (ms) | 1,176 (630) | 1,089 (484) | 1,148 (492) |

## 3.1 Investigation of the distress induction on subjective, physiological and behavioral components

To perform this investigation, we compared the features obtained from the Difficult Game (DG) condition with the Difficult Stressful Game (DSG) condition.

Regarding the subjective component, the global mental workload score, computed from the NASA-TLX, was significantly higher for DSG than for DG ($W(29) = 20$, $p < .001$, $r = -.533$). This effect has been also found on the Frustration/Stress subscale of the same questionnaire ($W(29) = 47$, $p < .001$, $r = -.449$). We also investigated the subjective component using the global ratings of negative feelings on the emotional wheel and found an effect of the DGS condition ($t(29) = -2.934$, $p = .003$, $d = -0.545$). These findings are summarized in Figure 7.

Regarding heart rate, we found an effect of the DSG condition compared to the DG condition ($W(26) = 66$, $p = .002$, $r = -.386$). This effect was also found on the respiration rate ($W(22) = 53$, $p = .008$, $r = -.360$) and on the number of SCRs ($t(21) = -2.863$, $p = .005$, $d = -0.625$). There was no difference regarding the amplitude of SCRs, nor on the tonic mean amplitude. We present those findings in Figure 8.

At the behavioral level, an effect of the DSG on the number of alarm omissions ($W(28) = 24.5$, $p = .035$, $r = -.254$) was found. There were 2 omissions in DG condition, while 8 alarms were omitted in DSG condition. Note that all these omissions were distributed on 8 participants only. There were no differences regarding performance in the game.

### 3.2 Investigation of biofeedback regulation on physiological entrainment and impacts on subjective and behavioral components

In order to test our hypothesis regarding the impact of the regulation on the physiological component, we compared the physiological features computed from DSG vs those from Difficult Stressful Game with Regulation (DSGR) condition. These analyses failed to evidence any entrainment phenomenon on physiology that could have been elicited by the biofeedback (Figure 8). However, we observed a reduction of HR for some participants (N=13) between DSG vs DSGR, yielding to -7.4 bpm ($M = -2.413$). The reduction was significant for these 13 participants, for DSG vs the DSGR ($t = 3.731$, $p_{Bonferroni} < .01$).

Regarding the subjective features, there was no effect of the biofeedback on the NASA-TLX scores ($p = .220$, $d = 0.233$), neither on the negative feelings scores of the GEW ($p = .096$, $d = 0.320$). These are presented on Figure 7.

Exploratory analysis using repeated measures ANOVA, with the biofeedback perceived efficiency groups as between subject factor, showed that there was an interaction between the biofeedback perceived efficiency and negative feelings rating on GEW ($F(1,27) = 31.576$, $p_{Bonferroni} < .0001$, $\eta^2 = .079$). Post hoc analysis suggests that, within "efficient" group ($N = 12$), there is a significant difference of negative feelings between DSG ($M = 2.767$, $SD = 0.660$) and DSGR ($M = 1.967$, $SD = 0.743$; $t = 5.906$, $p_{Bonferroni} < .0001$). However, no differences were found regarding physiological parameters associated with these groups, thus reinforcing the conclusion that there was no entrainment effect of the regulation on physiology, even when the biofeedback was subjectively efficient.

Regarding the behavioral component, no differences were found on the game performance indicator, on the total number of alarm omissions nor on reaction times, between DSG and DSGR. It should be noted that, amongst the 4 participants with at least one omission in DSG, and who rated the biofeedback as efficient, 3 of them did not miss any alarm during DSGR. Within DSGR, participants with omissions were mainly those who evaluated the biofeedback as "non-efficient".

**Figure 7**

*Results obtained for the 3 conditions (DG, DSG, DSGR) for the subjective component for (a) NASA-TLX and (b) negative feelings on Geneva Emotional Wheel. Regarding p-value, \* indicates a p-value < .05, while \*\* indicates a p-value <.01, NS=Non-significant. Error bars reflects the 95% confidence interval.*

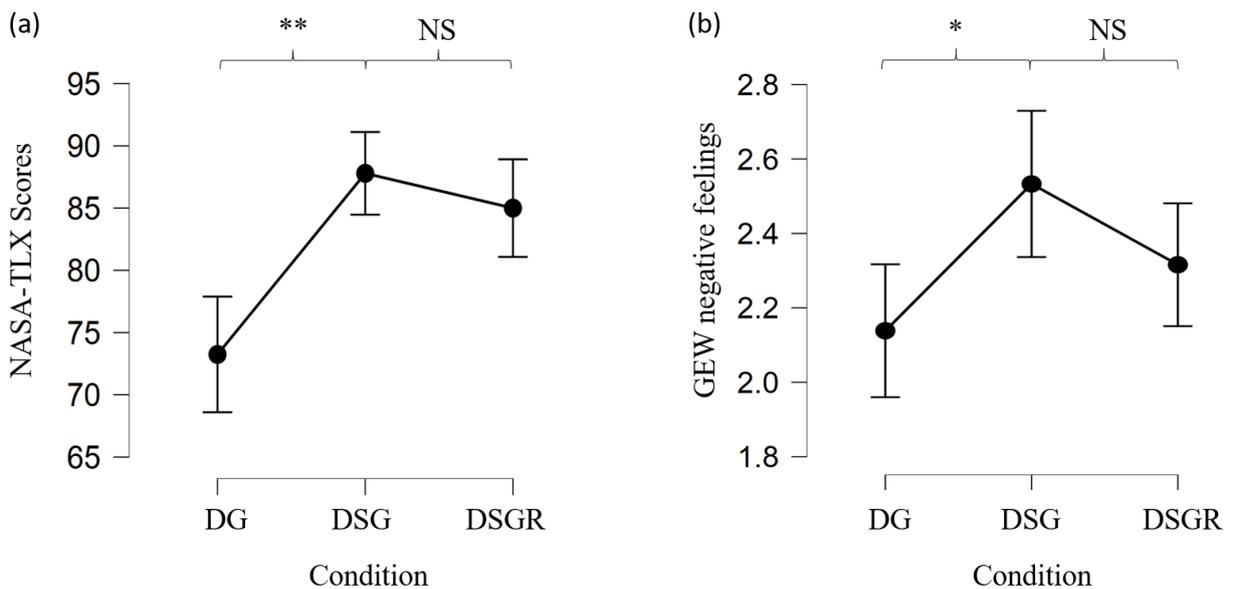

**Figure 8**

*Results obtained for the 3 conditions (DG, DSG, DSGR) for the baseline-normalized physiological component, with (a) heart rate, (b) respiration rate, and (c) number of skin conductance responses. Regarding p-value, * indicates a p-value < .05, NS=Non-significant. Error bars reflects the 95% confidence interval*

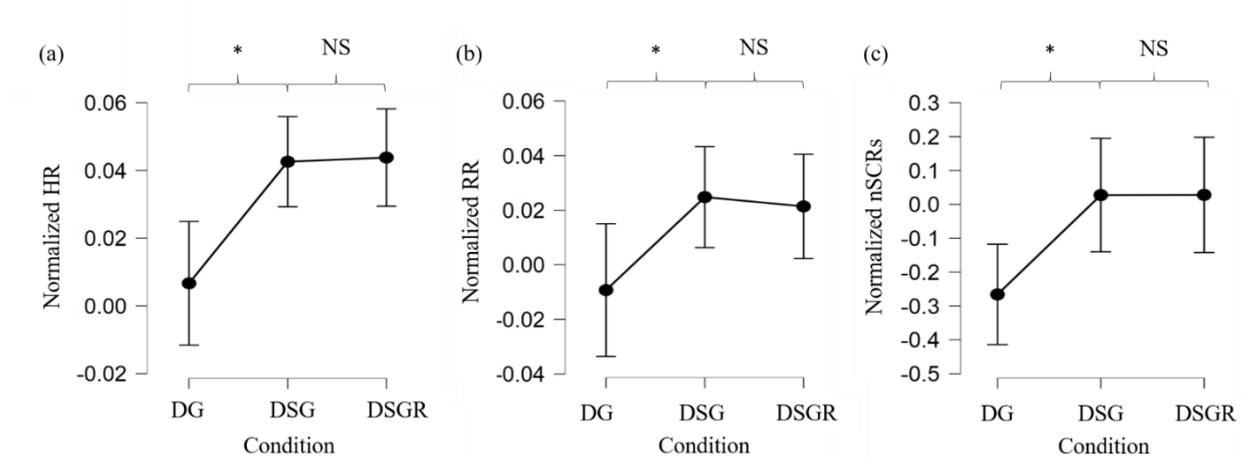

### 3.3 Investigation of the impacts of individual traits on biofeedback efficiency

The investigation of individual traits linked with biofeedback efficiency was performed through correlation analysis between personality traits and the variation between DSG and DSGR for the features representing the 3 components of stress response. We report significant correlations in Table 3.

**Table 3**

*Significant correlations between individual traits and features variation between DSGR and DSG*

| Personality trait | Variable | Spearman's $r$ | $p$ value |
|---|---|---|---|
| Extraversion | Negative feelings on GEW DSGR-DSG | -.409 | .027 |
| Neuroticism | NASA-TLX DSGR-DSG | -.523 | .004 |
| Openness | nSCRs normalized DSGR-DSG | .517 | .014 |

We formed extraversion and neuroticism groups (2 groups, High and Low for each trait according the median value). According to the correlation analysis results, we were expecting a significant reduction of the negative feelings between DSG and DSGR for the High extraversion group. A one-way ANOVA with the extraversion group as between subject factors revealed an interaction between the condition and the extraversion groups on GEW negative feelings ($F(1,27) = 9.047$, $p = .006$, $\eta^2 = .037$). Planned contrasts, according to our hypothesis, showed a significant difference between DSG ($M = 2.897$, $SD = 0.634$) and DSGR ($M = 2.357$, $SD = 1.001$) within the high extraversion group ($t(27) = -3.497$, $p = .002$). We also found a significant difference between the 2 groups within DSG ($t(34.568) = -2.399$, $p = .022$). Figure 9$_a$ shows the GEW negative feelings scores for extraversion groups.

On the basis of the correlation analysis, we were expecting a similar pattern regarding the NASA TLX score for neuroticism. A one-way ANOVA with the neuroticism group as between subject factors revealed an interaction between the condition and the neuroticism groups on the NASA TLX ($F(1,27) = 6.864$, $p = .014$, $\eta^2 = 0.034$). Planned contrasts, according to our hypothesis, showed a significant difference between DSG ($M = 90.71$, $SD = 12.63$) and DSGR ($M = 82.42$, $SD = 15.57$) within the high neuroticism group ($t(27) = 2.842$; $p = .008$). Figure 9$_b$ shows the NASA-TLX scores for neuroticism groups.

There were no significant results for the other features for both traits.

**Figure 9**

*Subjective results for a) extraversion groups, and b) neuroticism groups, between Difficult Stressful Game and Difficult Stressful Game with Regulation. Regarding p-value, \* indicates a p-value < .05, NS=Non-significant. Error bars reflects the 95% confidence interval*

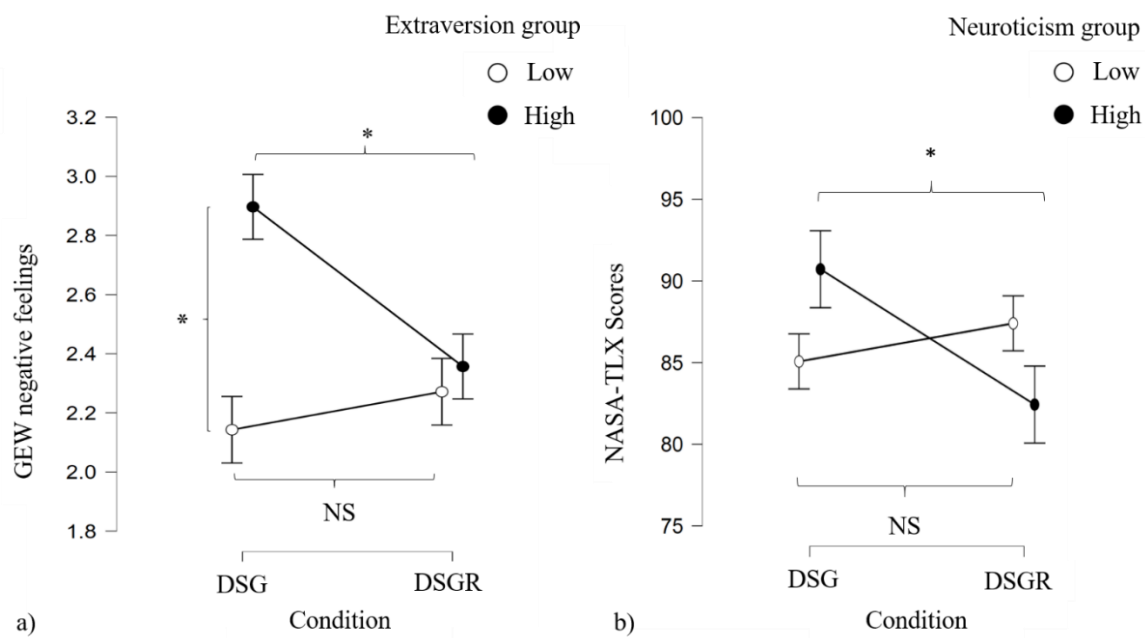

# 4     DISCUSSION

This work investigated the efficiency of a haptic stress regulation technique. This regulation took the form of reduced rate heartbeat-like vibrations depending on the participant's real time heart rate. After checking the protocol successfully induced distress, this study had 3 main goals:

-Firstly, to verify whether a reduced real-time biofeedback (BF) of heart rate could elicit an entrainment effect on the physiological component of the stress response.

- Secondly, to check if this BF could reduce the feeling of distress and its deleterious behavioral effects.

-Finally, the third objective of the study was linked to the exploration of individual traits that could impact the efficiency of the BF.

We found the expected variations of the stress metrics during DSG, corroborating the literature. Results showed an increased arousal during this condition, reflected by an increase in the number of skin conductance responses, an augmentation of heart and respiration rates. We also found subjective impacts of the protocol, translating into increased negative feelings during DSG and increased evaluation of the mental load induced by the condition. Regarding the behavioral component, we measured a significant increase of alarm omissions in DSG which may be related to an attentional tunneling effect induced by stress. When occurring within the auditory modality, this effect has been coined in the literature as "Inattentional Deafness" (ID, Dehais et al., 2019).

However, some discrepancy exists in the results on DSG. Firstly, the behavioral effect of stress was mostly on the auditory detection task and not on the game performance. A possible explanation for this could lie in the difference between "effectiveness" and "efficiency". The first term refers to the actual measurable performance, while the second term refers to the amount of effort mobilized to achieve the performance (Eysenck et al., 2007). Hence, in the game task, to obtain an equal performance (i.e. equal effectiveness) participants would have had to mobilize higher levels of cognitive resources in DSG, than in DG to manage the stressors (i.e. decreased efficiency). This would have led to an increased subjective mental load and distress, and to the observed physiological changes. As the paradigm was a dual task where the stress level was manipulated by the game, participants may have prioritized this performance at the expense of the auditory task, leading to the behavioral effects on this monitoring activity. Another contrast in our results was the absence of effects on heart rate variability or on the amplitude of SCRs. This might be explained by the difference in sensitivity of these parameters that is often noted in the literature.

As for the investigation of the BF impact on physiology, our hypothesis was that the presence of the real-time reduced BF could reduce the arousal (i.e. due to an entrainment effect). This study is, to our knowledge, the first one to explore the impact of a real-time reduction of the tactile BF stimulation, synchronized with the actual heart rate. While some participants had a significant reduction of their heart rate in DSGR, our results did not allow us to draw the conclusion that a global physiological entrainment occurred: no significant modulation of the various features computed to investigate the physiological component (cardiac and respiratory activity and skin-conductance responses) was found during DSGR.

Regarding the second objective, we did not obtain global variations on the subjective nor on the behavioral component. However, concerning the behavioral component, it should be noted that BF seems to have had an impact on the attentional tunneling induced by the stressful condition: within DSGR, participants with omissions were mainly those who evaluated BF as "non-efficient". This suggests that BF may have helped some participants with the management the auditory task, while distracting others. Yet, given the small number of participants with omissions, we could not backup this statement with appropriate analyses. Focusing on the subjective component, while our results cannot allow us to conclude that BF did have a global subjective impact on the participants, a significant correlation between extraversion and the subjective negative feelings was found. A significant correlation was also found for neuroticism and subjective mental load.

In line with the third objective, the High extraversion group had a stronger negative reaction than the Low-trait group toward stressful stimuli in DSG. It was found that BF allowed a stronger decrease in negative feelings for this group in DSGR compared to DSG. Regarding neuroticism, the High-trait group presented a significant decrease in the perceived mental load in DSGR compared to DSG, but not in negative feelings.

Our interpretation of these results is that our BF implementation elicited a subtle attentional deployment toward the vibrations, based on peripheral attention (Bakker & Niemantsverdriet, 2016). Indeed, we did not find behavioral changes regarding game performance: in case of perturbation from the BF, we would have expected a degradation of this performance. This interpretation could also be supported by the fact that, as mentioned in introductory section 1.4, extraverted people tend to have a facilitating attentional bias toward positive stimuli. The tactile biofeedback could thus have been perceived as a positive stimulation, leading to a decrease in feelings associated with distress for the people scoring high on extraversion scale. However, we found that people with extraversion seem to have higher negative feelings when confronted to DSG. As literature points out, given the beneficial effects of extraversion on coping during stressful situations, we would have expected the opposite effect (fewer negative feelings for the highest extraverted of our sample in DGS). A possible explanation for this could be linked to the observation that people scoring higher in extraversion trait tend to develop problem-focused coping (Penley & Tomaka, 2002). Nonetheless, in the present protocol, participants could not use a coping strategy based on a modulation of the situational demand, as the difficulty of the game and the temporal pressure was adaptative to the performance. It is therefore possible that these design constraints would have elicited further frustration for the most extraverted people, as the preferred problem-focused coping would have been inefficient and the control level low, similar to the study from LeBlanc, Ducharme & Thompson (2004). Such frustration could only have been alleviated when a positive stimulation was added in the situation, such as BF stimulation.

Our interpretation regarding our results on neuroticism is quite similar: due to the distraction from BF, people exhibiting a high neurotic trait in our sample would have partly shifted their attention away from the stressful stimuli on which they tended to focus on, thus eliciting a greater distress reduction for this group than for the Low neurotic group. This idea, that distraction-based strategies could thus be efficient for high neurotic traits, has already been suggested in the literature (Kobylińska et al., 2019). Furthermore, neurotic people tend to be pessimistic regarding emotion regulation (Purnamaningsih, 2017), and to relate negatively to emotion regulation strategies (John & Gross, 2007). Henceforth, people who had a higher score of neuroticism may have had a greater propensity to interpret their change of state as a mental load decrease rather than a negative feeling decrease.

Thus, **it seems to be not only crucial to consider personality aspects when evaluating stress regulation, but also the initial level of stress to regulate**. Indeed, these traits would have an impact not only on the emergence of stress, but also on the haptic regulation in our study. The mechanisms involved for these impacts seem to be different depending on the dominant trait. This might explain the discrepancies in previous results from the literature on haptic regulation. Moreover, as noted in the literature, the effect size of the haptic BF seems quite small, which also makes the question of the sensitivity of the measures important to consider. **This underlines the relevance of having a multidimensional evaluation of not only the stress response, but also of the individual, when working on regulation technologies**. The use of machine learning algorithms on physiological data could help to disentangle various personality traits (Evin et al., 2022). Our results are complementary to those from Xu et al. (2021), who studied the influence of interoceptive abilities on the physiological effect of false HR haptic regulation devices. However, contrary to our results, they failed to evidence subjective effect while evidencing physiological effects. We did not study the interoceptive abilities of our sample, but on the basis of their results and ours, we can point out **that inter-individual variations have a great influence not only on which Gross' regulation is taking place, but also on the efficiency of this regulation**.

This study has several limitations. The main limitation lies in the number of participants and the difficulty to precisely investigate individual traits within such a restrained sample size. Regarding this limitation, the suggestions we make, while being supported by literature, should be considered carefully. Moreover, in our study, the BF was deceitful as the participants were informed that the tactile stimulation would reflect their actual HR. As previous studies found that BF could be efficient even when participants are aware that the stimulations are lowered, it would have been interesting to further explore the impact of such knowledge on BF efficiency. Similarly, regarding the differences in subjective evaluation during the experimentation, an alternative interpretation could be that the participants would have (consciously or not) considered the rhythm delivered during the DSGR condition when evaluating their state *a posteriori*. This raises concerns over the influence of manipulation checks on the state of the participants. Such limitations have already been identified by Hauser et al. (2018). In the same line, the participants were carrying a lot of measuring instruments. Gržinič Frelih et al. (2017), showed that the use of body sensors may affect the participant's perception of the experiment and the collected data. While participants had a baseline measurement, wore the equipment for a certain period of time before the beginning of the experimentation (for the training phase, etc.), and despite the fact that the conditions were counterbalanced, it is possible that participants have been affected by these manipulation checks.

From an applied standpoint, this study reveals some **points of vigilance regarding peripheral monitoring activities**, such as those linked to computerized interactions. They will likely become more accessible to large audiences during the coming years, especially with the advent of technologies such as automated driving, and could include distressful stimulations. Hence, **further investigation of the impact of wearable regulatory methods within such contexts could be pertinent** as it could bring safety guidelines regarding the use of new technologies, but could also increase the comfort level of their users. **Other regulation techniques based on a more explicit response modulation may be interesting to explore within those contexts**, such as respiratory-based regulation methods, as previously mentioned. Studies show that they may have strong impacts on physiological entrainment and subsequent subjective modulation, while being suitable in environments demanding a certain level of sustained attention such as driving, especially with the use of haptics (Paredes et al., 2018).

## 5    CONCLUSION

This paper investigated the impacts of a stress regulation method based on a real-time lowered HR biofeedback using haptic stimulations. The main findings are that there was a moderate effect of this strategy on the subjective feeling of stress, but depending on the extraversion and neuroticism level, and on the level of stress to regulate. We found that while this effect was not attributable to an entrainment effect of the physiological activity, it may be linked to an attentional deployment toward the haptic stimulation, as found by previous studies in the auditory modality. These findings illustrate the complex interactions that exist between stress, emotion regulation and individual factors. These interactions should be considered when investigating haptic wearables designed to perform stress regulation. This study also allowed us to provide perspectives regarding the impacts of stress on attentional tunneling during peripheral monitoring activities, and the implementation of efficient stress regulation interactive wearables within such contexts.

## 6    CONFLICT OF INTEREST

The authors declare that the research was conducted in the absence of any commercial or financial relationships that could be construed as a potential conflict of interest.

## 7    ACKNOWLEDGMENTS

The authors would like to thanks Bertrand Richard, Bruno Piechnik and Romain Derollepot for their insights and help regarding the device used to design the game, but also Dr. Liam Elliott for proofreading the manuscript.

## 8     FUNDING

The work benefited from the support of University Gustave Eiffel, as part of the PhD of A.J. Béquet.